\documentclass[twocolumn,showpacs,amsmath,amssymb,prl,superscriptaddress]{revtex4}

\usepackage{graphicx}
\usepackage{dcolumn}
\usepackage{bm}
\usepackage{color} 

\begin{document}


\title{Momentum Distribution and Renormalization Factor in Sodium and the Electron Gas}

\author{Simo Huotari} 
\affiliation{European Synchrotron Radiation Facility, B.P. 220, F-38043 Grenoble, France}
\affiliation{Department of Physics, P. O. Box 64, FI-00014, University of Helsinki, Finland}
\author{J.\ Aleksi Soininen}
\affiliation{Department of Physics, P. O. Box 64, FI-00014, University of Helsinki, Finland}
\author{Tuomas Pylkk{\"a}nen}
\affiliation{European Synchrotron Radiation Facility, B.P. 220, F-38043 Grenoble, France}
\affiliation{Department of Physics, P. O. Box 64, FI-00014, University of Helsinki, Finland}
\author{Keijo H\"am\"al\"ainen}
\affiliation{Department of Physics, P. O. Box 64, FI-00014, University of Helsinki, Finland}
\author{Arezki Issolah}
\affiliation{Universit{\'e} de Tizi-Ouzou, Campus de Hasnaoua, 15000 Tizi-Ouzou, Algeria}
\author{Andrey Titov}
\affiliation{Institut N\'eel, CNRS \& UJF, F-38042 Grenoble, France}
\author{Jeremy McMinis}
\affiliation{Dept. of Physics and NCSA, U. of Illinois at
Urbana-Champaign, Urbana, IL 61801, USA}
\author{Jeongnim Kim}
\affiliation{Dept. of Physics and NCSA, U. of Illinois at
Urbana-Champaign, Urbana, IL 61801, USA}
\author{Ken Esler}
\affiliation{Dept. of Physics and NCSA, U. of Illinois at
Urbana-Champaign, Urbana, IL 61801, USA}
\author{David M. Ceperley}
\affiliation{Dept. of Physics and NCSA, U. of Illinois at
Urbana-Champaign, Urbana, IL 61801, USA}
\author{Markus Holzmann}
\affiliation{LPTMC, UPMC-CNRS, Paris, France, and LPMMC, UJF-CNRS, F-38042 Grenoble, France}
\author{Valerio Olevano}
\affiliation{Institut N\'eel, CNRS \& UJF, F-38042 Grenoble, France}

\date{\today}

\begin{abstract}
We present experimental and theoretical results on the momentum
distribution and the quasiparticle renormalization factor in sodium.
From an x-ray Compton-profile measurement of the valence-electron
momentum density, we derive its discontinuity at the Fermi wavevector.
This yields an accurate measure of the renormalization factor that we
compare with quantum Monte Carlo and $G_0W_0$ calculations performed
both on crystalline sodium and on the homogeneous electron gas.  Our
calculated results are in good agreement with the experiment.
\end{abstract}


\maketitle

\paragraph{Introduction} 

The homogeneous electron gas (HEG), also known as jellium, is one of
the most fundamental models in condensed matter physics \cite{electrongas}.
It is one of the simplest many-body systems which can still
describe several properties of real solids, especially of
the alkali metals.  For almost half a century, the
accurate description of many-body correlation effects has challenged
quantum many-body theory
and HEG is the canonical
workbench to test different theoretical methods
\cite{hedin65,rice65,pajanne82,lam71,holm98,lantto80,takada91}.
Although the exact analytic solution of the many-body problem in HEG is
still unknown,
today quantum Monte Carlo (QMC) calculations are widely accepted
to provide the most
reliable results on e.g.\ the correlation energy.
The situation is less clear concerning
spectroscopic quantities such as the momentum distribution, $n(p)$.
The accuracy of the theoretical methods in this respect
is not well understood, different approaches yielding a wide range of
varying results. This fundamental issue has remained unresolved,
mainly due to a lack of accurate, bulk-sensitive 
and unambiguous experimental probes that could be used to compare the theories
with.

Experimentally, one of nature's closest realizations of HEG is formed
by the valence electrons in alkali metals, especially Na. 
Here, we present very
accurate experimental and theoretical results on the electron-momentum
distribution of Na.  The single occupied valence band of Na has an
almost spherical Fermi surface and its properties in ambient
conditions with a density parameter $r_s=3.99$ 
can be directly compared with theoretical results on HEG.  In particular, we obtain
a precise experimental reference value for the 
quasiparticle
renormalization factor, $Z_{k_F}$, which characterizes the
discontinuity of the momentum distribution at the Fermi surface at
this density \cite{electrongas}. 

From the Compton profile (CP) measured by inelastic x-ray scattering
experiments on bulk sodium, we derive $n(p)$ and obtain
$Z^\textrm{Na}_{k_F}=0.58(7)$.  Compared to previous experiments
\cite{eisenberger72}, our experimental resolution provides a clear
observation of the discontinuity at the Fermi surface in a direct and
model-independent way.

We compare our experimental results to theoretical calculations using
QMC and $G_0W_0$ methods, both done for HEG and for solid Na taking
into account the electron-electron interaction and band-structure
effects.  Our calculations confirm the jellium-like behavior of Na and
allow us to quantify the small band-structure-induced 
deviations from HEG.
Finally, we compare the results with other many-body
approximations applied to HEG in literature
\cite{hedin65,rice65,pajanne82,lam71,holm98,lantto80,takada91}
(Table~\ref{table:zfs}).
Unless explicitly specified,  we use atomic units (a.u.).

\begin{figure}
\includegraphics[width=0.8\linewidth]{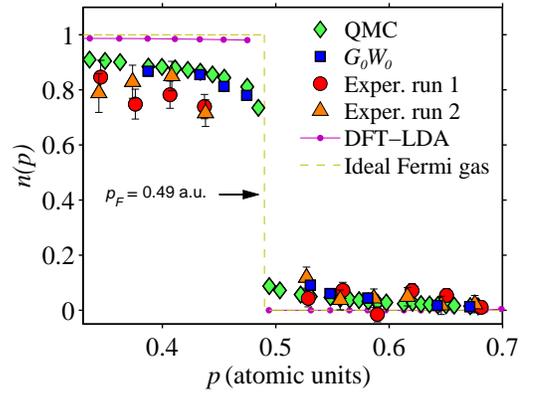}
\caption{\label{figu:nofp} (Color online) 
The  momentum distribution
of Na determined by
experiment, QMC using 
Slater-Jastrow 
wavefunctions, $G_0W_0$, 
and DFT-LDA calculations. The ideal-Fermi gas step function is also shown.
}
\end{figure}

\begin{table}
\begin{tabular}{llll}
\hline 
\hline
Technique & $\zeta^\text{Na}$ & $Z^\textrm{Na}_{k_F}$ &  $Z^\textrm{HEG}_{k_F}$\\
\hline
Experiment &   0.57(7) &   0.58(7) \\
QMC-SJ   & 0.68(2)  & 0.70(2) & 0.69(1) \\
QMC-BF &  & &0.66(2)  \\
$G_0W_0$ & 0.64(1) & 0.65(1) & 0.64 \cite{hedin65} \\
$GW$   & & & 0.793 \cite{holm98} \\
RPA (on-shell)  &  & & 0.45 \cite{pajanne82,lam71}\\
$\exp S_2$  &  & & 0.59 \cite{pajanne82}\\
EPX  &  & & 0.61 \cite{takada91} \\
Lam   &  & & 0.615 \cite{lam71}\\
FHNC  &  & & 0.71 \cite{lantto80}\\
\hline
\hline
\end{tabular}
\caption{\label{table:zfs} 
Our results for $\zeta$ and $Z_{k_F}$ of Na and HEG at $r_s=4.0$:
experimental value compared to $G_0W_0$, as well as QMC 
results based on Slater-Jastrow (QMC-SJ) and
more precise backflow (QMC-BF)
wavefunctions.
Various theoretical values of HEG from literature are also quoted.
}
\end{table}

\paragraph{Theory}
The momentum distribution per spin state, $n(p)$ (see Fig.\ 1),
is one of the basic many-body
observables where the Pauli principle for fermions is directly visible.  
It is the probability to observe an electron with 
momentum $p$.  For a 
non-interacting Fermi gas at zero temperature, 
$n(p)$ is 1  for $p$ below the Fermi
momentum $p_F$ and 0 above, i.e. 
$n(p) = \theta(p_F-p)$ with a discontinuity $\zeta = n(p_F^-) - n(p_F^+) = 1$ 
occurring at the Fermi surface.  
For a non-interacting crystalline
system, the electrons occupy Bloch wavefunctions,
 $\phi_{\nu\mathbf{k}}(\mathbf{r}) = \sum_\mathbf{G} \tilde{\phi}^{\mathbf{G}}_{\nu \mathbf{k}}
e^{i(\mathbf{k}+\mathbf{G})\mathbf{r}}$, 
where $\mathbf{k}$ is the crystal momentum,
$\nu$ the band index, and
 $\mathbf{G}$ are
reciprocal lattice
vectors. 
For systems like Na with one valence band ($\nu=1$)
whose Fermi surface is entirely contained within the first Brillouin
zone (1BZ), 
the band structure reduces the discontinuity 
of
the momentum distribution,
$\zeta=|\tilde{\phi}^{\mathbf{G}=0}_{\nu=1,\mathbf{k}_F}|^2 < 1$.
From a density-functional theory  
calculation within the local-density approximation (DFT-LDA) 
for Na, we obtain $\zeta^{\text{Na}}_{\text{DFT}}=0.98(1)$.
The  calculated valence band is an almost
perfect parabola, its wavefunction is nearly isotropic, and its Fermi
surface deviates from a perfect sphere by only 0.2\%.
These deviations of the Fermi surface from a perfect sphere necessarily
lead to a further, albeit small, reduction of the discontinuity
when $n(p)$ is orientationally averaged.

Many-body effects introduce a much larger reduction of the
discontinuity at the Fermi surface. This is known as the quasiparticle
renormalization factor, $Z_{\mathbf{k}_F}$.
Theoretical predictions for $Z_{\mathbf{k}_F}$ using different approximations range from $0.45$ to
$0.79$ for the density considered (see Table I).
In general,
the renormalization factor $Z_{\mathbf{k}_F}$ is related to 
the self-energy $\Sigma_{\nu,\nu}(k,\omega)$ via
$Z_{\mathbf{k}_F} = (1 - \partial \Sigma_{1,1}(\mathbf{k}_F,\omega) /
\partial \omega |_{\omega={\epsilon_F}})^{-1}$,  
$\Sigma=\Sigma^\textrm{e-e}+\Sigma^\textrm{e-p}$ 
containing
 the electron-electron
interactions,  $\Sigma^\textrm{e-e}$, and the
electron-phonon effects, $\Sigma^\textrm{e-p}$.
For HEG the discontinuity in the momentum
distribution is $\zeta_{\text{HEG}}=Z_{k_F}$.
In a jellium-like system such as Na, 
band-structure effects and many-body correlations 
can be factorized
so that
$\zeta_{\text{Na}}=|\tilde{\phi}^{\mathbf{G}=0}_{\nu=1,\mathbf{k}_F}|^2 Z_{\mathbf{k}_F}$ 
with a renormalization factor very close to the value for HEG
at the same density, if phonon effects can be neglected.

To determine $Z_{\mathbf{k}_F}$ theoretically, 
we performed pseudopotential 
diffusion QMC \cite{QMC-review,filippi99} calculations of bulk sodium
 based on a Slater-Jastrow (SJ) wavefunction using the QMCPACK code,
 and more precise calculations using backflow (BF) for HEG.
Complementary to QMC, we have done a non-self-consistent
(one-shot) $G_0W_0$ calculation \cite{hedin65}
starting from the DFT-LDA electronic structure
using the ABINIT code.

Within both methods, pseudopotentials are used to 
describe the core electrons,
based on a regular
static lattice for the ions, neglecting effects due to 
electron-phonon coupling.
Whereas core correlation effects only give smooth corrections that
do not influence the value of the renormalization factor,
electron-phonon coupling may lead to a further decrease of $Z_{\mathbf{k}_F}$. 
However, since the phonon Debye frequency $\omega_D$ is small compared
to the Fermi energy,  main effects of $\Sigma^\textrm{e-p}$ are expected only
 within a  narrow
momentum region around $p_F$, with $\delta p/p_F \lesssim 
\omega_D/ p_F^2 \approx 10^{-2}$. 
As we will see below, those effects
are beyond the resolution of the
experiment. The static approximation and the use of pseudopotential
should thus be sufficient to obtain the 
value of $Z_{\mathbf{k}_F}$,
whereas they may be less accurate to predict the  whole CP.

\paragraph{Experiment}

A unique bulk-sensitive probe of the momentum density is offered by
Compton scattering of x-rays \cite{cooperbook}.  The
experiment measures the spectra of x-rays scattered by an electron
system.  When the energy transferred to the electron is much larger
than its binding energy, the so-called impulse approximation (IA)
is valid and the measured spectrum is
related to the CP, which in isotropic average normalized to one electron is
\begin{equation}
\label{equ:CP}
J(q) = \frac{3}{8 \pi p_F^3} \int_{4 \pi} \mathrm{d}\Omega \int_{|q|}^\infty  p~ n(\mathbf{p})~\mathrm{d}p.
\end{equation}
Here $q$ is the component of the ground-state momentum of the electron
projected onto the scattering vector. 
Assuming an isotropic system, $n(p)$ can thus be
extracted by a differentiation of the CP,
\begin{equation}
\label{equ:np-from-CP}
n(p) = -\frac{2p_F^3}{3p} \frac{\mathrm{d}J(q)}{\mathrm{d}q}\Big |_{q=p}.
\end{equation}
For the non-interacting
HEG the CP is an inverted parabola $J(q) = \frac{3}{4 p_F^3}(p_F^2-q^2)$ 
for $q < p_F$ and vanishes for $q>p_F$  ($p_F=0.49(1)$~a.u.\ for Na). 
Many-body effects promote a part of the electrons from below to above
$p_F$. The CP, while being always continuous,
should retain a kink, i.e.\ a discontinuity in the first derivative
at $k_F$. Measuring the CP accurately 
allows the extraction of $n(p)$ by using
Eq.\ (\ref{equ:np-from-CP}).
The determination of $Z_{k_F}$
via x-ray Compton scattering has been a long-standing goal of many
scientists \cite{cooperbook}. 
The simultaneous requirements of extremely high momentum resolution
and statistical accuracy as well as difficulties in separating
band-structure and correlation effects have made such attempts difficult,
leading to anomalously low values of e.g.\ $Z_{k_F}^\mathrm{Li}=0.1(1)$ 
for Li ($r_s=3.25$) 
\cite{schulke96}. 
Promising results were given for Al ($r_s=2.07$) 
\cite{suortti00} by comparisons with 
an analytical model of $n(p)$ 
with an adjustable $Z_{k_F}$ \cite{schulke96}, giving the 
best agreement with $Z^\mathrm{Al}_{k_F} \approx 0.7$. This
determination however assumed a specific shape of $n(p)$ and thus was not
model-independent. Our choice of a HEG-like system of Na combined with
ultra-high resolution measurements allows us to accurately 
determine the $n(p)$ and
$Z_{k_F}^\mathrm{Na}$ in a model-independent way.

The experiments were performed at the beamline ID16 \cite{verbeni09}
of the European Synchrotron Radiation Facility on polycrystalline
sodium.  
The spectrometer was based on
a Rowland circle with  spherically bent analyzer
crystals with a Bragg angle of 89$^\circ$.  The
measurements were done by changing the incident-photon energy $E_1$
and observing the flux of scattered photons into a fixed scattering
angle $2\theta$ at a fixed energy $E_2$. 
We used
two different configurations and
photon-energy ranges to verify the result in two independent ways.  In
the first experiment ({\em run~1}), we used a 
single Si(555) analyzer crystal, $E_2=9.9$~keV, $2\theta=147^\circ$
and $E_1=9.9$--11.0~keV. In the second experiment ({\em run~2}), two
Si(880) analyzers were used, with $E_2=12.9$~keV, $2\theta=149^\circ$
and $E_1=12.9$--14.0~keV. 
The sample
was prepared in a glove box and transported to the beamline within an
argon atmosphere and pumped into a vacuum of $10^{-6}$~mbar. There was
no observable degradation of the sample when it was inspected after
the experiment.  
The measured signal was corrected for 
sample self-absorption as well as changes
in the incident photon flux and the spectra were measured repeatedly
to identify any possible instabilities during the experiment. None
were found and the spectra were finally averaged.

\begin{figure}
\includegraphics[width=0.9\linewidth]{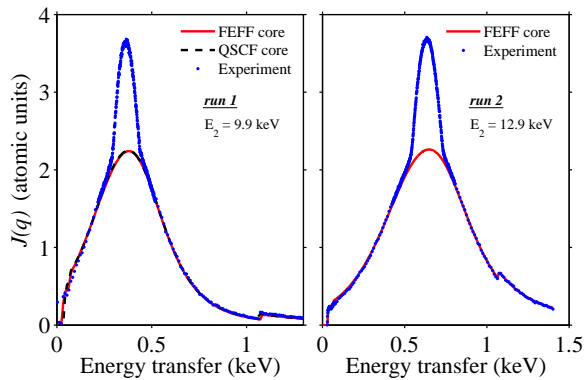}
\caption{(Color online) \label{figu:na_exp} 
The measured x-ray-scattering spectra from Na as a 
function of energy transfer, for both experimental runs. 
The experimental spectra consist of overlapping
valence and core contributions. Theoretical core contributions are shown 
for both QSCF and FEFFq treatments.}
\end{figure}

The measured spectra as a function of energy transfer are shown in
Fig.\ \ref{figu:na_exp}. The Na $L$ edges are seen at 30--60 eV, and
$K$ edge at 1.07 keV. In {\em run~1} the CP is centered at an energy
transfer of 365~eV and in {\em run~2} at 645~eV. 
Since our interest is in the valence-electron CP, the core contribution
has to be subtracted first. Since the IA is not strictly valid 
for the core-electron spectra in these experiments, we calculated them
with two independent methods: using i) the quasi-self-consistent field
(QSCF) approximation \cite{issolah88} and ii) the real-space multiple
scattering approach with the FEFF program \cite{feff} with
modifications for calculating the momentum-dependent scattering
cross-section \cite{feffq}. 
The differences between the two approaches are negligible. 
The core contribution can then be reliably 
subtracted from the experimental spectra.
The spectra can now 
be converted into the CP \cite{holm88}; for each energy transfer we
can evaluate the scattering-electron momentum component $q$ and the
measured intensity is related to the probability of finding the
electron with that $q$.

A finite experimental accuracy in the determination of $q$ will
introduce a broadening of any sharp features in the experimental data.
This uncertainty is caused in the present experiments by the spread of
scattering angles of the detected radiation.  This geometrical
contribution to the $q$-resolution was $\Delta q=0.018$~a.u.\
({\em run 1}) and $\Delta q=0.027$~a.u.\ ({\em run 2})
full-width-at-half-maximum (FWHM).  Final-state effects
\cite{soininen01,sternemann00}, i.e.\ the interaction of the
scattering electron and the rest of the electron gas, are known to
cause further broadening of the measured valence Compton profiles. We
calculated the magnitude of this broadening \cite{soininen01}, and
found it to be effectively an additional Gaussian smoothing of
0.08~a.u.\ ({\em run 1}), and 0.03~a.u.\ ({\em run 2}) (FWHM). This
combined with the geometrical resolutions yields effective
experimental $q$-resolutions of 0.08~a.u.\ ({\em run 1}) and 0.04~a.u.\
({\em run 2}).

\begin{figure}
\includegraphics[width=0.8\columnwidth]{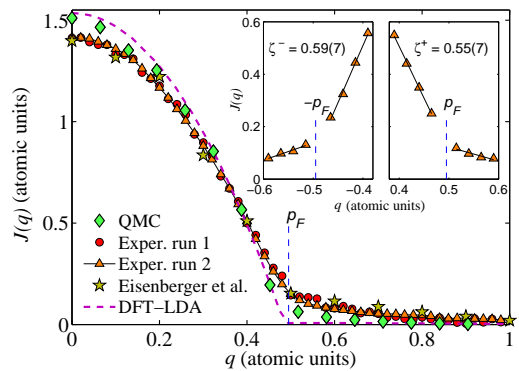}
\caption{\label{Compton}(Color online) 
The experimentally determined valence CPs
from the two runs (averaged over $q<0$ and $q>0$), 
compared to the results of Eisenberger {\em et al.} 
\cite{eisenberger72}, and to the QMC CP in the
(100) direction.  Inset: zoom on the discontinuity (points) and the
linear fits to the CP around the Fermi momentum. Both runs give
the same result within the errorbars, the fits shown here are
from {\em run 2}. }
\end{figure}

\paragraph{Results and discussion}
The result of the experiment, after the analysis described above, is the
valence CP shown in Fig.\ \ref{Compton}.  The valence CP of a
real metal in general deviates from the jellium parabola due to two
reasons: (i) correlation modifies the $n(p)$ introducing tails for
$p>p_F$, and (ii) electron-ion interaction modifies the overall
wavefunction and induces tails for $p>p_F$ due to
core-orthogonalization and the high-momentum components 
$\tilde{\phi}^{\mathbf{G} \neq 0}_{\nu \mathbf{k}}$.  
As discussed above, the valence
electron wavefunction of Na is fully contained inside 1BZ and is
highly free-electron-like,
with negligible high-momentum components.
The band structure only leads to small ($\lesssim 3\%$)
lowering of the momentum distribution for $p<p_F$, as can be seen in the
difference between the ideal-Fermi-gas and the DFT-LDA results in Fig.\ 1.
For this reason, we can compare the experimental 
momentum distribution and the CP to those of HEG
after taking these small corrections into account.

In Fig.\ \ref{Compton}, the Fermi momentum can be directly seen as the
discontinuity of the valence CP derivative. From the experimental data
we deduce $p_F=0.49(1)$~a.u.\ (LDA value $0.481$~a.u.).  The best
determination of the magnitude of the discontinuity at the Fermi
surface is provided by linear fits to the measured points in the
immediate vicinity of $p_F$.  An inset to Fig.\ \ref{Compton} shows
these fits, here for both negative and positive $q$ from 
the higher-$q$-resolution {\em run 2}.
The difference of the slopes for the two sides gives $\zeta^{-} =
0.59(7)$ and $\zeta^{+} = 0.55(7)$, which allows us to quote the
average value as $\zeta^\mathrm{Na}_\mathrm{exp} = 0.57(7)$. The
errorbar is based on the statistical noise of $J(q)$ 
and the uncertainty of $p_F$ (cf.\ Eq.\ (2)). 
Using the pure
band-structure value in LDA,
$|\tilde{\phi}^{\mathbf{G}=0}_{\nu=1,\mathbf{k}_F}|^2=0.98$,
we deduce the experimental
$Z_{k_F}=\zeta^\mathrm{Na}_\mathrm{exp}/|\tilde{\phi}^{\mathbf{G}=0}_{\nu=1,\mathbf{k}_F}|^2=0.58(7)$. 
We simulated the effect of 
the finite experimental $q$-resolution by
convoluting the QMC CP with the resolution function 
of {\em run 2}, and repeating the analysis described above. The 
result was an effective lowering of $\zeta_\mathrm{QMC}^\mathrm{Na}$ by 0.02.
Since this effect is smaller than our errorbar of 0.07, no further correction
to the final experimental result was found necessary.

The momentum distribution can be calculated from the 
experimental CP using Eq.~(2)
and is shown in Fig.\ 1. In the differentiation, the effect of statistical 
noise increases and thus we only 
show the result after averaging the values of $p<0$ and $p>0$, 
as well as averaging adjacent measured points. Due to this, the 
quantitative determination of $\zeta$ is better done by directly
analyzing the CP as described above. However, the trend of Table I
is clearly seen also in Fig.\ 1 and the experimental $n(p)$
is consistent with the obtained $\zeta^\text{Na}_\text{exp}$.

We have also calculated the CP by QMC in the (100)
direction (Fig.\ \ref{Compton}); small differences with respect to the
experiment remain.  Further calculations are necessary to study if the
inclusion of core electrons or phonon effects are necessary to reduce
the residual discrepancy between theory and experiment in the CP.  In
Fig.\ \ref{figu:nofp}, we show the resulting direction-averaged
$n(p)$. Finite-size corrections \cite{holzmann09} are important to
determine the structure around $p_F$, in particular to obtain the jump
at the Fermi surface, $\zeta^{\text{Na}}_{\text{QMC}}=0.68(2)$.
Within the QMC approach, we can determine bare band-structure effects
by turning off the explicit electron-electron correlations in the
underlying many-body wavefunction which yields a band-structure
contribution $|\tilde{\phi}^{\mathbf{G}=0}_{\nu=1,\mathbf{k}_F}|^2 =0.97(1)$
compatible with that of DFT-LDA.  We therefore obtain
$Z_{k_F}^\mathrm{Na}=0.70(2)$, in agreement with QMC calculations of HEG
using the same type of wavefunctions (SJ).  Within HEG, we have
performed calculations using more accurate Slater-Jastrow backflow
(BF) wavefunctions which indicate a slightly lower value,
$Z_{k_F}^\mathrm{HEG}=0.66(2)$.

From the $G_0W_0$ self-energy, we directly obtain 
$Z_{\mathbf{k}_F} = (1 - \partial \Sigma_{1,1}(\mathbf{k}_F,\omega) /
\partial \omega |_{\omega={\epsilon_F}})^{-1} = 0.65(1)$.
In order to determine the jump in $n(p)$, we further need the
quasi-particle weight which, within $G_0W_0$, coincides with the
Kohn-Sham orbital of the DFT-LDA calculation at the Fermi energy,
and we get $\zeta^\mathrm{Na}_{G_0W_0}=0.64(1)$. The momentum distribution
within $G_0W_0$ is very close to the QMC result.

The agreement between QMC-BF and $G_0W_0$ is remarkable.
Furthermore,
within both theories band-structure effects can be factorized 
from correlations within the accuracy of the
calculation, which enables a direct comparison with HEG.
In Table I, we summarize our experimental and theoretical results on 
$Z_{\mathbf{k}_F}$ and compare them
 to various
theoretical results obtained for HEG. 
On theoretical grounds, 
QMC-BF is considered to give the most precise result.
Together with $G_0W_0$, it is
in reasonable  agreement with the experiment.

Our experimental and theoretical values
clearly exclude two different classes
of approximation, and thus resolve 
a long-standing theoretical controversy.
Whereas the so-called on-shell approximation of the RPA
 \cite{pajanne82,lam71}
leads to an underestimation of $Z_{\mathbf{k}_F}$,
fully self-consistent $GW$ calculations
 \cite{holm98} overestimate its value. 
The latter result shows in particular
that the use of the
theoretically more appealing conserving approximation ($GW$)
does not result in an improved description of  spectral quantities
compared to non-self-consistent ($G_0W_0$) treatments.

\paragraph{Conclusions} 
We have determined the 
momentum distribution of Na valence
electrons experimentally and theoretically.  In particular, we have
related the discontinuity at the Fermi level to the quasiparticle 
renormalization
factor of HEG at $r_s=3.99$, giving a long-sought reference value 
for this fundamental quantity. 

Beamtime was provided by the ESRF. The authors would like to thank
Gy.\ Vank\'o, G.\ Monaco, R.\ Verbeni, H.\ M{\"u}ller 
and C.\ Henriquet (ESRF) for expert advice and assistance.
J.A.S., T.P.\ and K.H. were supported by the Academy of Finland contract
No.\ 1127462.  
Computer time was provided by the DOE Incite allocation. 
Funding was provided by Endstation.


\end{document}